\documentclass[10pt, conference, a4paper, oneside, twocolumn]{IEEEtran}
\usepackage{graphicx}
\usepackage{amsmath}
\usepackage{amssymb}
\usepackage{mathrsfs}
\usepackage{stfloats}
\usepackage{dsfont}
\usepackage{setspace}
\begin{document}
%
%
%
%\input COVER.tex
%\newpage
%
%
%
\title{On the Performance of Space Shift Keying (SSK) Modulation with Imperfect Channel Knowledge}
\author{\authorblockN{Marco Di Renzo$^{(1)}$, Dario De Leonardis$^{(2)}$, Fabio Graziosi$^{(2)}$, Harald Haas$^{(3)}$}
\authorblockA{\footnotesize $^{(1)}$ L2S, UMR 8506 CNRS -- SUPELEC -- Univ Paris--Sud\\
Laboratory of Signals and Systems (L2S), French National Center for Scientific Research (CNRS)\\
\'Ecole Sup\'erieure d'\'Electricit\'e (SUP\'ELEC), University of Paris--Sud XI (UPS)\\
3 rue Joliot--Curie, 91192 Gif--sur--Yvette (Paris), France\\
$^{(2)}$ University of L'Aquila, College of Engineering\\
Department of Electrical and Information Engineering (DIEI), Center of Excellence of Research DEWS\\
Via G. Gronchi 18, Nucleo Industriale di Pile, 67100 L'Aquila, Italy\\
$^{(3)}$ The University of Edinburgh, College of Science and Engineering\\
School of Engineering, Institute for Digital Communications (IDCOM)\\
King's Buildings, Alexander Graham Bell Building, Mayfield Road, Edinburgh, EH9 3JL, UK\\
E--Mail: marco.direnzo@lss.supelec.fr, fabio.graziosi@univaq.it, h.haas@ed.ac.uk} \vspace{-0.5cm}}
\maketitle
\begin{abstract}
In this paper, we study the sensitivity and robustness of Space Shift Keying (SSK) modulation to imperfect channel knowledge at the receiver.
Unlike the common widespread belief, we show that SSK modulation is more robust to imperfect channel knowledge than other state--of--the--art
transmission technologies, and only few training pilots are needed to get reliable enough channel estimates for data detection. More precisely,
we focus our attention on the so--called Time--Orthogonal--Signal--Design (TOSD--) SSK modulation scheme, which is an improved version of SSK
modulation offering transmit--diversity gains, and provide the following contributions: i) we develop a closed--form analytical framework to
compute the Average Bit Error Probability (ABEP) of a mismatched detector for TOSD--SSK modulation, which can be used for arbitrary
transmit--antenna, receive--antenna, channel fading, and training pilots; ii) we perform a comparative study of the performance of TOSD--SSK
modulation and the Alamouti code under the same imperfect channel knowledge, and show that TOSD--SSK modulation is more robust to channel
estimation errors; iii) we point out that only few pilot pulses are required to get performance very close to the perfect channel knowledge
lower--bound; and iv) we verify that transmit-- and receive--diversity gains of TOSD--SSK modulation are preserved even for a mismatched
receiver.
\end{abstract}
%
%
%\IEEEpeerreviewmaketitle
%
%
%
\section{Introduction} \label{Intro}
\PARstart{S}{pace} modulation is a novel digital modulation concept for Multiple--Input--Multiple--Output (MIMO) wireless systems, which is
receiving a growing attention due to the possibility of realizing low--complexity and spectrally--efficient MIMO implementations
\cite{Yang_2008}--\cite{MDR_TCOM2010}. The space modulation principle is known in the literature in different forms, such as
Information--Guided Channel Hopping (IGCH) \cite{Yang_2008}, Spatial Modulation (SM) \cite{Haas_TVT}, and Space Shift Keying (SSK) modulation
\cite{Ghrayeb_TWC}. Although different from one another, all these transmission technologies share the same fundamental working principle,
which makes them different from conventional modulation schemes: they encode part of the information bits into the spatial position of the
antenna--array, which plays the role of a constellation diagram (the so--called ``spatial--constellation diagram'') for data modulation
\cite{MDR_TCOM2010}. In particular, SSK modulation exploits only the spatial--constellation diagram for data modulation, which results in a
very low--complexity modulation concept for MIMO systems \cite{Ghrayeb_TWC}. Recently, we have introduced in \cite{MDR_TOSD},
\cite{MDR_TCOM_TOSD} and generalized in \cite{MDR_TOSD_GSSK}, respectively, an improved version of SSK modulation, which can achieve
transmit--diversity gains without any spectral efficiency loss with respect to the original SSK modulation proposal.

In SSK modulation, blocks of information bits are mapped into the index of a single transmit--antenna, which is switched on for data
transmission while all the other antennas radiate no power \cite{Ghrayeb_TWC}. Regardless of the information message to be transmitted, and,
thus, the active transmit--antenna, SSK modulation exploits the location--specific property of the wireless channel for data modulation
\cite{MDR_TCOM2010}: the messages sent by the transmitter can be decoded at the destination since the receiver sees a different channel impulse
response on any transmit--to--receive wireless link. In \cite{Ghrayeb_TWC} and \cite{MDR_TCOM2010}, it has been shown that the achievable
performance of SSK modulation depends on how different the channel impulse responses are. In other words, the channel impulse responses are the
points of the spatial--constellation diagram, and the receiver performance depends on the distance among these points. Recent results have
shown that SSK modulation can provide better performance than conventional modulation schemes with similar complexity if the receiver has
Perfect Channel State Information (P--CSI) \cite{Yang_2008}--\cite{Ghrayeb_TWC}. However, due to its inherent working principle, the major
criticism about the application of SSK modulation in a realistic propagation environment is the robustness of the space modulation principle to
the imperfect knowledge, at the receiver, of the wireless channel. In particular, since the wireless channel is the actual modulation unit, it
is often argued that the space modulation concept is more sensitive to channel estimation errors. The main contribution of this paper is to
shed light on this matter.

Although some research works on the performance of the space modulation principle with imperfect channel knowledge are available in the
literature, these results are insufficient and only based on numerical simulations. In \cite{Ghrayeb_TWC}, the authors study the Average Bit
Error Probability (ABEP) of SSK modulation with non--ideal channel knowledge. However, there are four limitations in this paper: i) the ABEP is
obtained only through Monte Carlo simulations, which is not very much insightful; ii) the arguments in \cite{Ghrayeb_TWC} are applicable only
to Gaussian fading channels and do not take into account the cross--product between channel estimation error and Additive White Gaussian Noise
(AWGN) at the receiver; iii) it is unclear from \cite{Ghrayeb_TWC} how the ABEP changes with respect to the pilot symbols used by the channel
estimator; and iv) the robustness/weakness of SSK modulation with respect to conventional modulation schemes is not analyzed. In
\cite{MDR_TCOM_PCSI}, we have studied the performance of SSK modulation when the receiver does not exploit the knowledge of the phase of the
channel gains for data detection (semi--blind receiver). The main message of this paper is that semi--blind receivers are much worse than
coherent detection schemes, and, thus, that the assessment of the performance of coherent detection with imperfect channel knowledge is a
crucial aspect for SSK modulation. A very interesting study has been recently conducted in \cite{Ulla_Faiz}, where the authors have compared
the performance of SM and V--BLAST (Vertical Bell Laboratories Layered Space--Time) \cite{V_BLAST} schemes with practical channel estimates. It
is shown that the claimed sensitivity of space modulation to channel estimation errors is simply a misconception and that, on the contrary, SM
is more robust than V--BLAST to imperfections on the channel estimates, and that less training is, in general, required by SM. However, the
system in \cite{Ulla_Faiz} is studied only through Monte Carlo simulations, which does not give too much insights for performance analysis and
system optimization.

Motivated by these considerations, in this paper we aim at developing a very general analytical framework to assess the performance of the
space modulation concept with coherent detection and practical channel estimates. Our theoretical and numerical results corroborate the
findings in \cite{Ulla_Faiz}, and highlight two important outcomes: 1) space modulation can be even more robust to channel estimation errors
than conventional modulation schemes, and 2) the number of pilot symbols required to approach the lower--bound set by coherent detection with
perfect channel knowledge is quite limited. More precisely, the contributions of this paper are as follows: i) we develop a general analytical
framework to compute the ABEP of the TOSD--SSK modulation scheme with imperfect channel knowledge. The framework can handle arbitrary
transmit--antenna, receive--antenna, fading channel statistics, and number of pilot symbols used by the channel estimation unit. It is shown
that the mismatched detector of TOSD--SSK modulation \cite{Biglieri} can be cast in terms of a quadratic--form in complex Gaussian Random
Variables (RVs) when conditioning upon fading channel statistics, and that the ABEP can be computed by exploiting the Gil--Pelaez inversion
theorem \cite{MDR_QF}; ii) we compare the performance of TOSD--SSK modulation with the Alamouti scheme \cite{Alamouti}, which similar to
TOSD--SSK modulation can achieve transmit--diversity equal to two, and show that TOSD--SSK modulation is more robust to imperfect channel
knowledge; and iii) we show that transmit-- and receive--diversity of TOSD--SSK modulation with non--ideal channel estimates is always
preserved.

The reminder of this paper is organized as follows. In Section \ref{System_Model}, the system model is introduced and the TOSD--SSK modulation
scheme is briefly described. In Section \ref{Performance_Analysis}, the analytical framework to compute the ABEP with imperfect channel
knowledge is developed. In Section \ref{Results}, numerical results are shown to substantiate the main findings of the paper. Finally, Section
\ref{Conclusion} concludes the paper.
\section{System Model} \label{System_Model}
We consider a generic $N_t \times N_r$ MIMO system, with $N_t$ and $N_r$ being the number of transmit-- and receive--antenna, respectively.
TOSD--SSK modulation works as follows \cite{Ghrayeb_TWC}, \cite{MDR_TOSD}--\cite{MDR_TOSD_GSSK}: i) the transmitter encodes blocks of $\log _2
\left(  N_t \right)$ data bits into the index of a single transmit--antenna, which is switched on for data transmission while all the other
antennas are kept silent, and ii) the receiver solves a $N_t$--hypothesis detection problem to estimate the transmit--antenna that is not idle,
which results in the estimation of the unique sequence of bits emitted by the encoder. With respect to conventional SSK modulation
\cite{Ghrayeb_TWC}, in TOSD--SSK modulation the $i$--th transmit--antenna, when active, radiates a distinct pulse waveform ${w_i \left(  \cdot
\right)}$ for $i=1, 2,\ldots, N_t$, and the waveforms across the antennas are time--orthogonal, {\emph{i.e.}}\footnote{$\left(  \cdot  \right)^
*$ denotes complex--conjugate.}, $\int\nolimits_{ - \infty }^{ + \infty } {w_i \left( t \right)w_j^* \left( t \right)dt}  = 0$ if $i \ne j$ and
$\int\nolimits_{ - \infty }^{ + \infty } {w_i \left( t \right)w_j^* \left( t \right)dt}  = 1$ if $i = j$. We emphasize here that in TOSD--SSK
modulation a single antenna is active for data transmission and that the transmitted message is still encoded into the index of the
transmit--antenna and {\emph{not}} into the impulse response of the shaping filter. In other words, the proposed concept is different from
conventional Single--Input--Single--Output (SISO) schemes, which use Orthogonal Pulse Shape Modulation (O--PSM) \cite{TCOM_PSM} and are unable
to achieve transmit--diversity, as only a single wireless link is exploited for communication \cite{MDR_TCOM_TOSD}, \cite{MDR_TOSD_GSSK}. Also,
the TOSD--SSK modulation principle is different from conventional transmit--diversity schemes \cite{TD_CommMag}. Further details are available
in \cite{MDR_TCOM_TOSD} and \cite{MDR_TOSD_GSSK}, which are here omitted in order to avoid repetitions. In \cite{MDR_TCOM_TOSD},
\cite{MDR_TOSD_GSSK}, it is shown that the diversity gain of the TOSD--SSK modulation scheme is $2 N_r$, which results in a transmit--diversity
equal to two and a receive--diversity equal to $N_r$.

In this paper, the block of bits encoded into the index of the $i$--th transmit--antenna is called ``message'' and is denoted by $m_i$ for
$i=1, 2,\ldots, N_t$. The $N_t$ messages are equiprobable. Moreover, the related transmitted signal is denoted by $s_i \left( \cdot \right)$.
It is implicitly assumed in this notation that, if $m_i$ is transmitted, the analog signal $s_i \left( \cdot \right)$ is emitted by the $i$--th
transmit--antenna while the other antennas radiate no power.
\subsection{Notation} \label{Notation}
The main notation used in this paper is as follows. i) We adopt a complex--envelope signal representation. ii) $j = \sqrt { - 1}$ is the
imaginary unit. iii) $\left( {x \otimes y} \right)\left( t \right) = \int_{ - \infty }^{ + \infty } {x\left( \xi  \right)y\left( {t - \xi }
\right)d\xi }$ is the convolution of signals $x\left(  \cdot \right)$ and $y\left(  \cdot  \right)$. iv) $\left| {\cdot} \right|^2$ is the
square absolute value. v) $\textrm{E}\left\{ \cdot  \right\}$ is the expectation operator. vi) ${\mathop{\rm Re}\nolimits} \left\{ \cdot
\right\}$ and ${\mathop{\rm Im}\nolimits} \left\{ \cdot \right\}$ are the real and imaginary part operators, respectively. vii) $\Pr \left\{
\cdot \right\}$ denotes probability. viii) $Q\left( x \right) = \left( {{1 \mathord{\left/ {\vphantom {1 {\sqrt {2\pi } }}} \right.
\kern-\nulldelimiterspace} {\sqrt {2\pi } }}} \right)\int_x^{ + \infty } {\exp \left( { - {{t^2 } \mathord{\left/ {\vphantom {{t^2 } 2}}
\right. \kern-\nulldelimiterspace} 2}} \right)dt}$ is the Q--function. ix) $\hat m$ denotes the message estimated at the receiver. x) $E_m$ is
the average energy transmitted by each antenna that emits a non--zero signal. xi) $T_m$ denotes the signaling interval for each information
message $m_i$ ($i=1, 2, \ldots, N_t$). xii) The noise $\eta_l$ at the input of the $l$--th receive--antenna ($l=1,2,\ldots,N_r$) is a complex
AWGN process, with power spectral density $N_0$ per dimension. Across the receive--antenna, the noises $\eta_l$ are statistically independent.
xiii) $E_p$ and $N_p$ denote the energy transmitted for each pilot symbol and the number of pilot symbols used for channel estimation. xiv)
$\delta \left(  \cdot  \right)$ and $\delta _{ \cdot , \cdot }$ are the Dirac and Kronecker delta functions, respectively. xv) For ease of
notation, we set $\bar \gamma {\rm{ = }}{{E_m } \mathord{\left/ {\vphantom {{E_m } {\left( {N_0 } \right)}}} \right. \kern-\nulldelimiterspace}
{{N_0 }}}$ and $r_{pm} = {{E_p } \mathord{\left/ {\vphantom {{E_p } {E_m }}} \right. \kern-\nulldelimiterspace} {E_m }}$. xvi) $M_X \left( s
\right) = \textrm{E}\left\{ {\exp \left( { sX} \right)} \right\}$ and $\Psi _X \left( \nu  \right) = \textrm{E}\left\{ {\exp \left( { j\nu X}
\right)} \right\}$ denote Moment Generating Function (MGF) and Characteristic Function (CF) of RV X, respectively.
\subsection{Channel Model} \label{Channel_Model}
We consider a general frequency--flat slowly--varying channel model with generically correlated and non--identically distributed fading gains.
In particular ($i=1, 2, \ldots, N_t$, $l=1, 2, \ldots, N_r$):
\begin{itemize}
\item $h_{i,l} \left( t \right) = \alpha _{i,l} \delta \left( {t - \tau _{i,l} } \right)$ is the channel impulse response of the
transmit--to--receive wireless link from the $i$--th transmit--antenna to the $l$--th receive--antenna. $\alpha _{i,l}  = \beta _{i,l} \exp
\left( {j\varphi _{i,l} } \right)$ is the complex channel gain with $\beta _{i,l}$ and $\varphi _{i,l}$ denoting the channel envelope and
phase, respectively, and $\tau _{i,l}$ is the propagation time--delay.
\item The delays $\tau _{i,l}$ are assumed to be known at the receiver, {\emph{i.e.}}, perfect time--synchronization is considered.
Furthermore, we assume $\tau _{1,1} \cong \tau _{1,2} \cong \ldots \cong \tau _{N_t, N_r }$, which is a realistic assumption when the distance
between the transmitter and the receiver is much larger than the spacing between transmit-- and receive--antennas \cite{MDR_TCOM2010}. Due to
the these assumptions, the propagation delays can be neglected in the reminder of this paper.
\end{itemize}
\subsection{Channel Estimation} \label{Channel_Estimation}
Similar to \cite{Proakis_1968} and \cite{Gifford_2005}, we assume that channel estimation is performed by using a Maximum--Likelihood (ML)
detector and by observing $N_p$ pilot pulses that are transmitted before the modulated data. During the transmission of one block of
pilot--plus--data symbols, the wireless channel is assumed to be constant, {\emph{i.e.}}, a block--fading channel is considered. With these
assumptions, the estimates of the channel gains $\alpha _{i,l}$ ($i=1, 2, \ldots, N_t$, $l=1, 2, \ldots, N_r$) can be written as follows:
\begin{equation} \footnotesize
\label{Eq_1} \hat \alpha _{i,l}  = \hat \beta _{i,l} \exp \left( {j\hat \varphi _{i,l} } \right) = \alpha _{i,l}  + \varepsilon _{i,l}
\end{equation}
\noindent where $\hat \alpha _{i,l}$, $\hat \beta _{i,l}$, and ${\hat \varphi _{i,l} }$ are the estimates of $\alpha _{i,l}$, $\beta _{i,l}$,
and ${\varphi _{i,l} }$, respectively, at the output of the channel estimation unit, and $\varepsilon _{i,l}$ is the additive channel
estimation error, which can be shown to be complex Gaussian distributed with zero mean and variance $\sigma _\varepsilon ^2  = {{N_0 }
\mathord{\left/ {\vphantom {{N_0 } {\left( {E_p N_p } \right)}}} \right. \kern-\nulldelimiterspace} {\left( {E_p N_p } \right)}}$ per dimension
\cite{Proakis_1968}, \cite{Gifford_2005}. The channel estimation errors $\varepsilon _{i,l}$ are statistically independent and identically
distributed, as well as statistically independent of the channel gains and the AWGN at the receiver.
\begin{figure*}[!t]
\setcounter{equation}{9}
\begin{equation} \footnotesize
\label{Eq_10}
\begin{split}
 & {\rm{PEP}}\left( {{\rm{TX}}_{t_1 }  \to {\rm{TX}}_{t_2 } } \right) = \Pr \left\{ {\hat D_{\left. {t_1 } \right|m_{t_1 } }  < \hat D_{\left. {t_2 } \right|m_{t_1 } } } \right\} = \Pr \left\{ {\frac{{\hat D_{\left. {t_1 } \right|m_{t_1 } } }}{{N_0 }} < \frac{{\hat D_{\left. {t_2 } \right|m_{t_1 } } }}{{N_0 }}} \right\} \\
  &= \Pr \left\{ \begin{array}{c}
 \sum\limits_{l = 1}^{N_r } {\left[ {\frac{1}{2}\left( {\alpha _{t_1 ,l} \sqrt {\bar \gamma }  + \varepsilon _{t_1 ,l} \sqrt {\bar \gamma } } \right)^ *  \left( {\alpha _{t_1 ,l} \sqrt {\bar \gamma }  + \frac{{\tilde \eta_{t_1 ,l} }}{{\sqrt {N_0 } }}} \right) + \frac{1}{2}\left( {\alpha _{t_1 ,l} \sqrt {\bar \gamma }  + \varepsilon _{t_1 ,l} \sqrt {\bar \gamma } } \right)\left( {\alpha _{t_1 ,l} \sqrt {\bar \gamma }  + \frac{{\tilde \eta_{t_1 ,l} }}{{\sqrt {N_0 } }}} \right)^ *   - \frac{1}{2}\left| {\alpha _{t_1 ,l} \sqrt {\bar \gamma }  + \varepsilon _{t_1 ,l} \sqrt {\bar \gamma }} \right|^2 } \right]}  \\
  <  \\
 \sum\limits_{l = 1}^{N_r } {\left[ {\frac{1}{2}\left( {\alpha _{t_2 ,l} \sqrt {\bar \gamma }  + \varepsilon _{t_2 ,l} \sqrt {\bar \gamma } } \right)^ *  \frac{{\tilde \eta_{t_2 ,l} }}{{\sqrt {N_0 } }} + \frac{1}{2}\left( {\alpha _{t_2 ,l} \sqrt {\bar \gamma }  + \varepsilon _{t_2 ,l} \sqrt {\bar \gamma } } \right)\frac{{\tilde \eta_{t_2 ,l}^ *  }}{{\sqrt {N_0 } }} - \frac{1}{2}\left| {\alpha _{t_2 ,l} \sqrt {\bar \gamma }  + \varepsilon _{t_2 ,l} \sqrt {\bar \gamma } } \right|^2 } \right]}  \\
 \end{array} \right\} \\
 \end{split}
\end{equation}
\normalsize \hrulefill \vspace*{0pt}
\end{figure*}
\subsection{Mismatched ML--Optimum Detector} \label{ML_Detector}
In this paper, for data detection we consider the so--called mismatched ML--optimum receiver according to the definition given in
\cite{Biglieri}. In particular, a detector with mismatched metric estimates the complex channel gains as in (\ref{Eq_1}) and uses the result in
the same metric that would be applied if the channel were perfectly known. This detector can be obtained as follows.

Let $m_n$ with $n=1, 2,\ldots, N_t$ be the transmitted message\footnote{In order to avoid any confusion with the adopted notation, let us
emphasize that the subscript $n$ denotes the actual message that is transmitted, while the subscript $i$ denotes the (generic) $i$--th message
that is tested by the detector to solve the $N_t$--hypothesis detection problem. More specifically, for each signaling interval, $n$ is fixed,
while $i$ can take different values at the detector.}. The signal received after propagation through the wireless fading channel and impinging
upon the $l$--th receive--antenna can be written as follows:
\setcounter{equation}{1}
\begin{equation} \footnotesize
\label{Eq_2} r_l\left( t \right) = \tilde s_{n,l} \left( t \right) + \eta_l\left( t \right)\quad \quad {\rm{if}}\;m_n \;{\rm{is}}\;{\rm{sent}}
\end{equation}
\noindent where $\tilde s_{n,l} \left( t \right) = \left( {s_n  \otimes h_{n,l} } \right)\left( t \right) = \alpha _{n,l} s_n \left( {t}
\right) = \beta _{n,l} \exp \left( {j\varphi _{n,l} } \right)s_n \left( {t} \right)$ for $n=1, 2,\ldots, N_t$ and $l=1, 2,\ldots, N_r$.
Furthermore, in TOSD--SSK modulation we have $s_n \left( t \right) = \sqrt {E_m } w_n \left( t \right)$ for $n=1, 2,\ldots, N_t$.

In particular, (\ref{Eq_2}) is a general $N_t$--hypothesis detection problem  \cite[Sec. 7.1]{Simon}, \cite[Sec. 4.2, pp. 257]{VanTrees} in
AWGN, when conditioning upon fading channel statistics. Accordingly, the mismatched ML--optimum detector with imperfect CSI at the receiver is
as follows:
\begin{equation} \footnotesize
\label{Eq_3} \hat m = \mathop {\arg \max }\limits_{m_i \; {\rm{ for }} \; i = 1,2, \ldots ,N_t } \left\{ {\hat D_i } \right\}
\end{equation}
\noindent where $\hat D_i$ is the mismatched decision metric:
\begin{equation} \footnotesize
\label{Eq_4} \hat D_i  = \sum\limits_{l = 1}^{N_r } {\left[ {{\mathop{\rm Re}\limits} \left\{ {\int\nolimits_{T_m } {r_l \left( t \right)\hat
s_{i,l}^ *  \left( t \right)dt} } \right\} - \frac{1}{2}\int\nolimits_{T_m } {\hat s_{i,l} \left( t \right)\hat s_{i,l}^ *  \left( t \right)dt}
} \right]}
\end{equation}
\noindent and $\hat s_{i,l} \left( t \right) = \hat \alpha _{i,l} s_i \left( t \right) = \left( {\alpha _{i,l}  + \varepsilon _{i,l} }
\right)s_i \left( t \right)$ for $i=1, 2,\ldots, N_t$ and $l=1, 2,\ldots, N_r$. If the transmitted message is $m_n$, which results in switching
on the $n$--th transmit--antenna for data transmission, the detector will be successful in detecting the transmitted message, {\emph{i.e.}},
$\hat m = m_n$, if and only if $\mathop {\max }\limits_{i = 1,2, \ldots ,N_t } \left\{ {\hat D_i } \right\} = \hat D_n$.

By using (\ref{Eq_2}), the decision metric in (\ref{Eq_4}), when conditioning upon the transmission of message $m_n$, {\emph{i.e.}}, $\hat
D_{\left. i \right|m_n }$, can be written as follows ($n=1, 2,\ldots, N_t$, $i=1, 2,\ldots, N_t$):
\begin{equation} \footnotesize
\label{Eq_5}
 \hat D_{\left. i \right|m_n }  = \sum\limits_{l = 1}^{N_r } {{\mathop{\rm Re}\nolimits} \left\{ {\alpha _{n,l} \hat \alpha _{i,l}^ *  E_m \delta _{i,n}  + \hat \alpha _{i,l}^ *  \sqrt {E_m } \tilde \eta_{i,l} } \right\}}
  - \frac{{E_m }}{2}\sum\limits_{l = 1}^{N_r } {\left| {\hat \alpha _{i,l} } \right|^2 }
\end{equation}
\noindent with $\tilde \eta_{i,l}  = \int\nolimits_{T_m } {\eta_l \left( t \right)w_i^ *  \left( t \right)dt}$.
\section{Performance Analysis} \label{Performance_Analysis}
In this section, we summarize the main steps to compute the ABEP of the mismatched detector in (\ref{Eq_3}). To this end, we exploit the same
methodology proposed in \cite{MDR_TCOM2010} for a receiver with P--CSI, but generalize the derivation to account for channel estimation errors.
More specifically, the ABEP can be tightly upper--bounded as follows \cite[Eq. (34)]{MDR_TCOM2010}:
\setcounter{equation}{5}
\begin{equation} \footnotesize
\label{Eq_6} {\rm{ABEP}} \le \frac{1}{{2\left( {N_t  - 1} \right)}}\sum\limits_{t_1  = 1}^{N_t } {\sum\limits_{t_2  \ne t_1  = 1}^{N_t }
{{\rm{APEP}}\left( {{\rm{TX}}_{t_1 }  \to {\rm{TX}}_{t_2 } } \right)} }
\end{equation}
\noindent where ${\rm{APEP}}\left( {{\rm{TX}}_{t_1}  \to {\rm{TX}}_{t_2} } \right)$ denotes the Average\footnote{The expectation is here
computed over fading channel statistics.} Pairwise Error Probability (APEP) of the transmit--antenna ${{\rm{TX}}_{t_1} }$ and ${{\rm{TX}}_{t_2}
}$ with ${t_1},{t_2} = 1,2,\ldots,N_t$, {\emph{i.e.}}, the probability of detecting ${{\rm{TX}}_{t_2} }$ when, instead, ${{\rm{TX}}_{t_1} }$ is
actually transmitting. More specifically, ${\rm{APEP}}\left( {{\rm{TX}}_{t_1} \to {\rm{TX}}_{t_2} } \right)$ is the ABEP of an equivalent $2
\times N_r$ MIMO system where only the transmit--antenna ${{\rm{TX}}_{t_1} }$ and ${{\rm{TX}}_{t_2} }$ can be switched on for transmission. In
this section, exact closed--form expressions of the APEPs in (\ref{Eq_6}) are given.
\begin{figure*}[!t]
\setcounter{equation}{18}
\begin{equation} \footnotesize
\label{Eq_21} {\rm{PEP}}\left( {{\rm{TX}}_{t_1 }  \to {\rm{TX}}_{t_2 } } \right) = \frac{1}{2} - \frac{1}{\pi }\int\nolimits_0^{ + \infty }
{\frac{{{\mathop{\rm Im}\nolimits} \left\{ {\Upsilon \left( \nu  \right)\Upsilon \left( { - \nu } \right)\exp \left( {\Delta _{t_1 } \left( \nu
\right)\sum\limits_{l = 1}^{N_r } {\left| {\alpha _{t_1 ,l} } \right|^2 }  + \Delta _{t_2 } \left( { - \nu } \right)\sum\limits_{l = 1}^{N_r }
{\left| {\alpha _{t_2 ,l} } \right|^2 } } \right)} \right\}}}{\nu }d\nu }
\end{equation}
\normalsize \hrulefill \vspace*{0pt}
\end{figure*}
\subsection{Computation of PEPs} \label{PEP}
Let us start by computing the PEPs, {\emph{i.e.}}, the pairwise probabilities in (\ref{Eq_6}) when conditioning upon fading channel statistics.
From (\ref{Eq_3}), the ${{\rm{PEP}}\left( {{\rm{TX}}_{t_1 }  \to {\rm{TX}}_{t_2 } } \right)}$ is as follows:
\setcounter{equation}{6}
\begin{equation} \footnotesize
\label{Eq_7} {\rm{PEP}}\left( {{\rm{TX}}_{t_1 }  \to {\rm{TX}}_{t_2 } } \right) = \Pr \left\{ {\hat D_{\left. {t_1 } \right|m_{t_1 } }  < \hat
D_{\left. {t_2 } \right|m_{t_1 } } } \right\}
\end{equation}
\noindent where, from (\ref{Eq_5}), ${\hat D_{\left. {t_1 } \right|m_{t_1 } } }$ and ${\hat D_{\left. {t_2 } \right|m_{t_1 } } }$ can be
explicitly written as follows:
\begin{equation} \footnotesize
\label{Eq_8} \hat D_{\left. {t_1 } \right|m_{t_1 } }  = \sum\limits_{l = 1}^{N_r } {{\mathop{\rm Re}\nolimits} \left\{ {\hat \alpha _{t_1 ,l}^
*  \left( {\alpha _{t_1 ,l} E_m  + \sqrt {E_m } \tilde \eta_{t_1 ,l} } \right)} \right\}} - \frac{{E_m }}{2}\sum\limits_{l = 1}^{N_r } {\left|
{\hat \alpha _{t_1 ,l} } \right|^2 }
\end{equation}
\begin{equation} \footnotesize
\label{Eq_9} \hat D_{\left. {t_2 } \right|m_{t_1 } }  = \sum\limits_{l = 1}^{N_r } {{\mathop{\rm Re}\nolimits} \left\{ {\hat \alpha _{t_2 ,l}^
*  \sqrt {E_m } \tilde \eta_{t_2 ,l} } \right\}}  - \frac{{E_m }}{2}\sum\limits_{l = 1}^{N_r } {\left| {\hat \alpha _{t_2 ,l} } \right|^2 }
\end{equation}

From (\ref{Eq_8}) and (\ref{Eq_9}), the PEP in (\ref{Eq_7}) can be written as shown in (\ref{Eq_10}) on top of the next page, where we have: i)
used the identity ${\mathop{\rm Re}\nolimits} \left\{ {ab^ *  } \right\} = \left( {{1 \mathord{\left/ {\vphantom {1 2}} \right.
\kern-\nulldelimiterspace} 2}} \right)ab^ *   + \left( {{1 \mathord{\left/ {\vphantom {1 2}} \right. \kern-\nulldelimiterspace} 2}} \right)a^ *
b$, which holds for any pair of complex numbers $a$ and $b$; ii) normalized to $N_0$ both decision metrics in order to explicitly show the
Signal--to--Noise Ratio (SNR) $\bar \gamma  = {{E_m } \mathord{\left/ {\vphantom {{E_m } {N_0 }}} \right. \kern-\nulldelimiterspace} {N_0 }}$;
and iii) used (\ref{Eq_1}).

Let us now define:
\setcounter{equation}{10}
\begin{equation} \footnotesize
\label{Eq_11} \left\{ \begin{split}
 & d_{t_{1},l }  = A\left| {X_{t_{1},l } } \right|^2  + B\left| {Y_{t_{1},l } } \right|^2  + CX_{t_{1},l } Y_{t_{1},l }^ *   + C^ *  X_{t_{1},l }^ *  Y_{t_{1},l }  \\
 & d_{t_{2},l }  = A\left| {X_{t_{2},l } } \right|^2  + B\left| {Y_{t_{2},l } } \right|^2  + CX_{t_{2},l } Y_{t_{2},l }^ *   + C^ *  X_{t_{2},l }^ *  Y_{t_{2},l }  \\
 \end{split} \right.
\end{equation}
\noindent where $X_{t_{1,l} }  = \alpha _{t_1 ,l} \sqrt {\bar \gamma } + \varepsilon _{t_1 ,l} \sqrt {\bar \gamma }$, $Y_{t_{1,l} }  = \alpha
_{t_1 ,l} \sqrt {\bar \gamma }  + {{\tilde \eta_{t_1 ,l} } \mathord{\left/ {\vphantom {{\tilde \eta_{t_1 ,l} } {\sqrt {N_0 } }}} \right.
\kern-\nulldelimiterspace} {\sqrt {N_0 } }}$, $X_{t_{2,l} }  = \alpha _{t_2 ,l} \sqrt {\bar \gamma }  + \varepsilon _{t_2 ,l} \sqrt {\bar
\gamma }$, $Y_{t_{2,l} }  = {{\tilde \eta_{t_2 ,l} } \mathord{\left/ {\vphantom {{\tilde \eta_{t_2 ,l} } {\sqrt {N_0 } }}} \right.
\kern-\nulldelimiterspace} {\sqrt {N_0 } }}$, and $A = -{1 \mathord{\left/ {\vphantom {1 2}} \right. \kern-\nulldelimiterspace} 2}$, $B=0$, $C
= {1 \mathord{\left/ {\vphantom {1 2}} \right. \kern-\nulldelimiterspace} 2}$.

With these definitions, the PEP in (\ref{Eq_10}) can be simplified as follows:
\begin{equation} \footnotesize
\label{Eq_13} {\rm{PEP}}\left( {{\rm{TX}}_{t_1 }  \to {\rm{TX}}_{t_2 } } \right) = \Pr \left\{ {d_{t_1 }  - d_{t_2 }  < 0} \right\} = \Pr
\left\{ {d_{t_1 ,t_2 }  < 0} \right\}
\end{equation}
\noindent where $d_{t_1 }  = \sum\nolimits_{l = 1}^{N_r } {d_{t_1 ,l} }$, $d_{t_2 }  = \sum\nolimits_{l = 1}^{N_r } {d_{t_2 ,l} }$, $d_{t_1
,t_2 }  = d_{t_1 }  - d_{t_2 }$.

From \cite[Sec. III]{MDR_QF}, we note that the PEP in (\ref{Eq_13}) can be studied by exploiting the theory of ``quadratic--form'' receivers in
complex Gaussian RVs. More specifically, after a few algebraic manipulations, it can be shown that, when conditioning upon fading channel
statistics, $d_{t_1 }$ and $d_{t_2 }$ are two quadratic forms with CF equal to:
\begin{equation} \footnotesize
\label{Eq_14} \Psi _{d_t } \left( {\left. \nu  \right|{\boldsymbol{\alpha }}_t } \right) = \frac{{\left( {v_a v_b } \right)^{N_r } \exp \left\{
{\frac{{v_a v_b \left( { - \nu ^2 \gamma _{t,a}  + j\nu \gamma _{t,a} } \right)}}{{\left( {\nu  + jv_a } \right)\left( {\nu  - jv_b }
\right)}}} \right\}}}{{\left( {\nu  + jv_a } \right)^{N_r } \left( {\nu  - jv_b } \right)^{N_r } }}\end{equation}
\noindent where we have emphasized the conditioning upon all the fading channels $\alpha _{i,l}$ for $i=1,2,\ldots,N_t$ and $l=1,2,\ldots,N_r$
in the channel vector ${\boldsymbol{\alpha}_t }$ with $t \in \left\{ {t_1 ,t_2 } \right\}$, and have defined:
\begin{equation} \footnotesize
\label{Eq_15} \left\{ \begin{split}
 & \gamma _{t_1 ,a}  = \frac{{\bar \gamma }}{2}\left( {1 + \frac{1}{{N_p r_{pm} }}} \right)\sum\limits_{l = 1}^{N_r } {\left| {\alpha _{t_1 ,l} } \right|^2 }  = g_{t_1 ,a} \sum\limits_{l = 1}^{N_r } {\left| {\alpha _{t_1 ,l} } \right|^2 }  \\
 & \gamma _{t_1 ,b}  = \frac{{\bar \gamma }}{2}\sum\limits_{l = 1}^{N_r } {\left| {\alpha _{t_1 ,l} } \right|^2 }  = g_{t_1 ,b} \sum\limits_{l = 1}^{N_r } {\left| {\alpha _{t_1 ,l} } \right|^2 }  \\
 & \gamma _{t_2 ,a}  = \frac{{\bar \gamma }}{2}\sum\limits_{l = 1}^{N_r } {\left| {\alpha _{t_2 ,l} } \right|^2 }  = g_{t_2 ,a} \sum\limits_{l = 1}^{N_r } {\left| {\alpha _{t_2 ,l} } \right|^2 }  \\
 & \gamma _{t_2 ,b}  =  - \frac{{\bar \gamma }}{2}\sum\limits_{l = 1}^{N_r } {\left| {\alpha _{t_2 ,l} } \right|^2 }  = g_{t_2 ,b} \sum\limits_{l = 1}^{N_r } {\left| {\alpha _{t_2 ,l} } \right|^2 }  \\
 \end{split} \right.
\end{equation}
\noindent where $g_{t_1 ,a}  = \left( {{1 \mathord{\left/ {\vphantom {1 2}} \right. \kern-\nulldelimiterspace} 2}} \right)\bar \gamma \left( {1
+ {1 \mathord{\left/ {\vphantom {1 {\left( {N_p r_{pm} } \right)}}} \right. \kern-\nulldelimiterspace} {\left( {N_p r_{pm} } \right)}}}
\right)$, $g_{t_1 ,b}  = g_{t_2 ,a}  =  - g_{t_2 ,b}  = \left( {{1 \mathord{\left/ {\vphantom {1 2}} \right. \kern-\nulldelimiterspace} 2}}
\right)\bar \gamma$, and $v_a  = \sqrt {\left( {{1 \mathord{\left/ {\vphantom {1 4}} \right. \kern-\nulldelimiterspace} 4}} \right) + N_p
r_{pm} }  + \left( {{1 \mathord{\left/ {\vphantom {1 2}} \right. \kern-\nulldelimiterspace} 2}} \right)$, $v_b  = \sqrt {\left( {{1
\mathord{\left/ {\vphantom {1 4}} \right. \kern-\nulldelimiterspace} 4}} \right) + N_p r_{pm} }  - \left( {{1 \mathord{\left/ {\vphantom {1 2}}
\right. \kern-\nulldelimiterspace} 2}} \right)$.

By taking into account (\ref{Eq_15}), the CF in (\ref{Eq_14}) can be re--written in the very compact form as follows:
\begin{equation} \footnotesize
\label{Eq_17} \Psi _{d_t } \left( {\left. \nu  \right|{\boldsymbol{\alpha}_t }} \right) = \Upsilon \left( \nu  \right)\exp \left( {\Delta_t
\left( \nu \right)\sum\limits_{l = 1}^{N_r } {\left| {\alpha _{t,l} } \right|^2 } } \right)
\end{equation}
\noindent where $\Upsilon \left(  \cdot  \right)$ and $\Delta _t \left(  \cdot  \right)$ are independent of the fading channel gains, and are
defined as follows:
\begin{equation} \footnotesize
\label{Eq_18} \left\{ \begin{split}
 & \Upsilon \left( \nu  \right) = \frac{{\left( {v_a v_b } \right)^{N_r } }}{{\left( {\nu  + jv_a } \right)^{N_r } \left( {\nu  - jv_b } \right)^{N_r } }} \\
 & \Delta _t \left( \nu  \right) = \frac{{v_a v_b \left( { - \nu ^2 g_{t,a}  + j\nu g_{t,b} } \right)}}{{\left( {\nu  + jv_a } \right)\left( {\nu  - jv_b } \right)}} \\
 \end{split} \right.
\end{equation}

Let us emphasize that $d_{t_1 }$ and $d_{t_2 }$ are conditional quadratic forms in complex Gaussian RVs because both the receiver noise and the
channel estimation errors are complex Gaussian RVs. Moreover, since noises and channel estimation errors are statistically independent if $t_1
\ne t_2$, then it follows that $d_{t_1 }$ and $d_{t_2 }$ are two independent quadratic forms. As a consequence, the CF of $d_{t_1 ,t_2 }$ in
(\ref{Eq_13}) can be computed as:
\begin{equation} \footnotesize
\label{Eq_19} \Psi _{d_{t_1 ,t_2 } } \left( {\left. \nu  \right|{\boldsymbol{\alpha}_{t_1} },{\boldsymbol{\alpha}_{t_2} }} \right) = \Psi
_{d_{t_1 } } \left( {\left. \nu \right|{\boldsymbol{\alpha}_{t_1} }} \right)\Psi _{d_{t_2 } } \left( {\left. { - \nu }
\right|{\boldsymbol{\alpha}_{t_2} }} \right)
\end{equation}

From (\ref{Eq_17}) and (\ref{Eq_19}), the PEP in (\ref{Eq_13}) can be computed from \cite[Eq. (11)]{MDR_QF}, as follows:
\begin{equation} \footnotesize
\label{Eq_20}
\begin{split}
 &{\rm{PEP}}\left( {{\rm{TX}}_{t_1 }  \to {\rm{TX}}_{t_2 } } \right) = \frac{1}{2} - \frac{1}{\pi }\int\nolimits_0^{ + \infty } {\frac{{{\mathop{\rm Im}\nolimits} \left\{ {\Psi _{d_{t_1 ,t_2 } } \left( {\left. \nu  \right|{\boldsymbol{\alpha}_{t_1} },{\boldsymbol{\alpha}_{t_2} }} \right)} \right\}}}{\nu }d\nu }  \\
  &\hspace{1.5cm}= \frac{1}{2} - \frac{1}{\pi }\int\nolimits_0^{{\pi  \mathord{\left/
 {\vphantom {\pi  2}} \right.
 \kern-\nulldelimiterspace} 2}} {\frac{{{\mathop{\rm Im}\nolimits} \left\{ {\Psi _{d_{t_1 ,t_2 } } \left( {\left. {\tan \left( \xi  \right)} \right|{\boldsymbol{\alpha}_{t_1} },{\boldsymbol{\alpha}_{t_2} }} \right)} \right\}}}{{\sin \left( \xi  \right)\cos \left( \xi  \right)}}d\xi }  \\
 \end{split}
\end{equation}
\noindent which yields an exact, single--integral, and closed--form expression for analysis and design of TOSD--SSK modulation with channel
estimation errors.
\subsection{Computation of APEPs} \label{APEP}
Let us now remove the conditioning upon the fading channel in (\ref{Eq_20}). To this end, we first substitute (\ref{Eq_17}) and (\ref{Eq_19})
in (\ref{Eq_20}), as shown in (\ref{Eq_21}) on top of this page. Then, by averaging over the fading channels, we obtain:
\setcounter{equation}{19}
\begin{equation} \footnotesize
\label{Eq_22}
\begin{split}
 &{\rm{APEP}}\left( {{\rm{TX}}_{t_1 }  \to {\rm{TX}}_{t_2 } } \right) = \textrm{E}\left\{ {{\rm{PEP}}\left( {{\rm{TX}}_{t_1 }  \to {\rm{TX}}_{t_2 } } \right)} \right\} \\
  &\hspace{0.5cm}= \frac{1}{2} - \frac{1}{\pi }\int\nolimits_0^{ + \infty } {\frac{{{\mathop{\rm Im}\nolimits} \left\{ {\Upsilon \left( \nu  \right)\Upsilon \left( { - \nu } \right)M_{{\rm A}_{t_1 ,t_2 } \left( \nu  \right)} \left( 1 \right)} \right\}}}{\nu }d\nu }  \\
 \end{split}
\end{equation}
\noindent where we have introduced the RV ${\rm A}_{t_1 ,t_2 } \left( \nu  \right)$ as follows:
\begin{equation} \footnotesize
\label{Eq_23} {\rm A}_{t_1 ,t_2 } \left( \nu  \right) = \Delta _{t_1 } \left( \nu  \right)\sum\limits_{l = 1}^{N_r } {\left| {\alpha _{t_1 ,l}
} \right|^2 } + \Delta _{t_2 } \left( { - \nu } \right)\sum\limits_{l = 1}^{N_r } {\left| {\alpha _{t_2 ,l} } \right|^2 }
\end{equation}

In summary, (\ref{Eq_22}) provides an exact, single--integral, and closed--form expression of the APEP for a generic correlated and
non--identically distributed MIMO wireless channel. To compute (\ref{Eq_22}), only the MGF of RV ${\rm A}_{t_1 ,t_2 } \left( \nu \right)$ in
(\ref{Eq_23}) has to be known in closed--form. This MGF might be computed for a large variety of fading channel models as shown in
\cite{Simon}. As an example, let us consider the scenario in which all the wireless links are independent. In this case, $M_{{\rm A}_{t_1 ,t_2
} \left( \nu \right)} \left(  \cdot  \right)$ reduces to:
\begin{equation} \footnotesize
\label{Eq_24}
 M_{{\rm A}_{t_1 ,t_2 } \left( \nu  \right)} \left( s \right) = \prod\limits_{l = 1}^{N_r } {M_{\left| {\alpha _{t_1 ,l} } \right|^2 } \left( {s\Delta _{t_1 } \left( \nu  \right)} \right)}
 \times \prod\limits_{l = 1}^{N_r } {M_{\left| {\alpha _{t_2 ,l} } \right|^2 } \left( {s\Delta _{t_2 } \left( { - \nu } \right)} \right)}
\end{equation}
\noindent where the MGFs $M_{\left| {\alpha _{t,l} } \right|^2 } \left(  \cdot  \right)$ for $t \in \left\{ {t_1 ,t_2 } \right\}$ are available
in \cite{Simon} for many fading channel models.

From (\ref{Eq_24}) and \cite{Giannakis}, we can observe that the diversity achieved by the TOSD--SSK modulation scheme with a mismatched
receiver is the same as the diversity achieved with P--CSI, {\emph{i.e.}}, $2N_r$. We will confirm this finding in Section \ref{Results} with
the help of some numerical examples.
\section{Numerical and Simulation Results} \label{Results}
In this section, we show some numerical examples to study the performance of TOSD--SSK modulation in the presence of channel estimation errors
and compare it with the Alamouti scheme, which similar to TOSD--SSK modulation can offer a diversity gain equal to $2N_r$ \cite{Alamouti}. The
simulation setup used in our study is as follows: i) we consider independent Rayleigh fading with normalized unit power over all the wireless
links. The MGF needed to compute (\ref{Eq_24}) can be found in \cite[Eq. (2.8)]{Simon}; ii) in TOSD--SSK modulation the rate is qual to $R =
\log _2 \left( {N_t } \right)$; iii) as far the Alamouti scheme is concerned, we consider Multilevel Phase Shift Keying (M--PSK) modulation
with constellation size $M$ and rate $R = \log _2 \left( {M } \right)$; and iv) the orthogonal shaping filters needed in TOSD--SSK modulation
are obtained from Hermite polynomials \cite{TCOM_PSM}

\begin{figure}[!t]
\centering
\includegraphics [width=1.1\columnwidth] {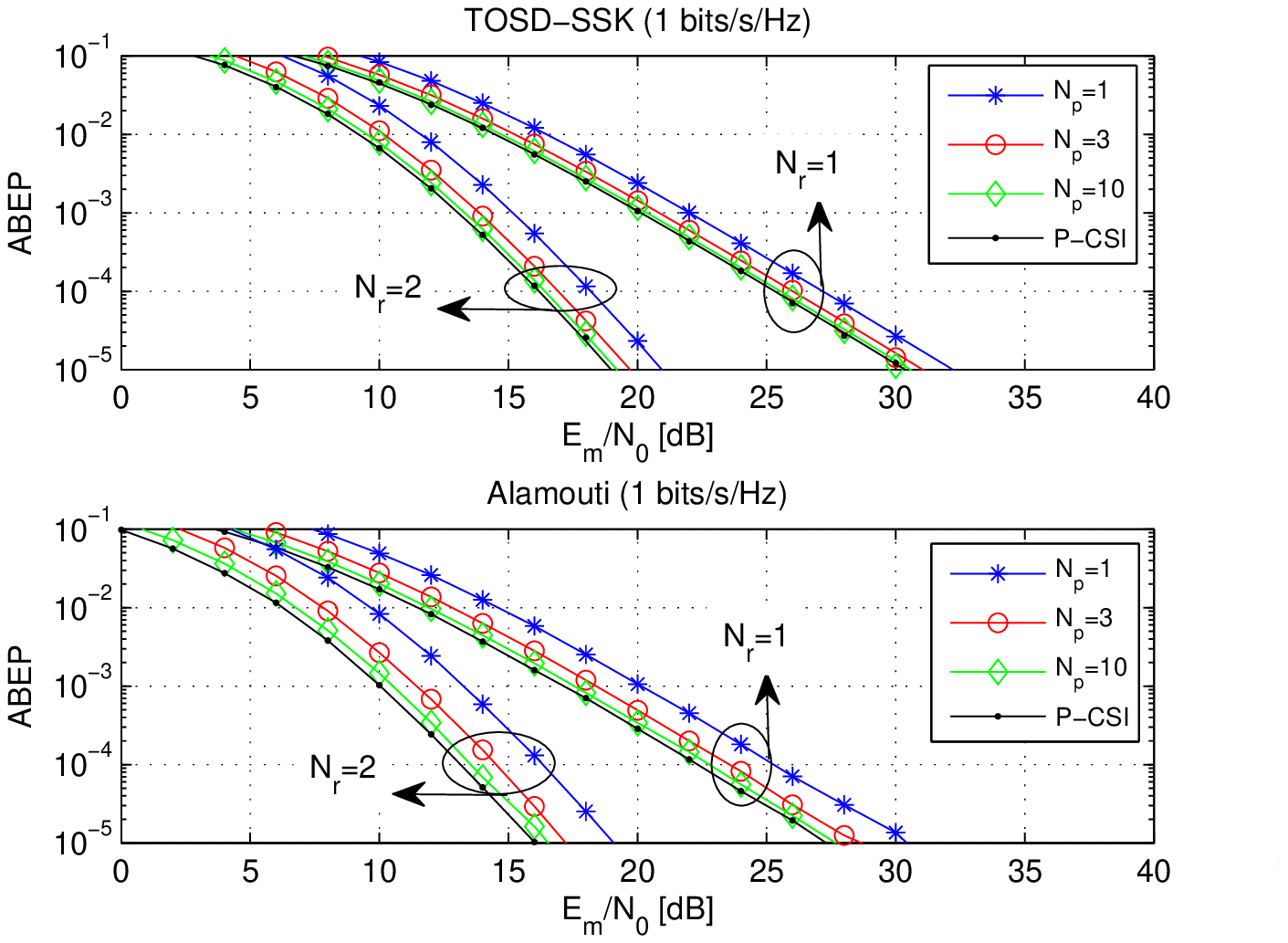}
\vspace{-0.5cm} \caption{\scriptsize ABEP against $E_m/N_0$ for various pilot pulses $N_p$ and rate 1 bits/s/Hz. (top) TOSD--SSK modulation:
solid lines show the analytical model and markers Monte Carlo simulations. (bottom) Alamouti scheme: only Monte Carlo simulations are shown.}
\label{Fig_R1}
\end{figure}

\begin{figure}[!t]
\centering
\includegraphics [width=1.1\columnwidth] {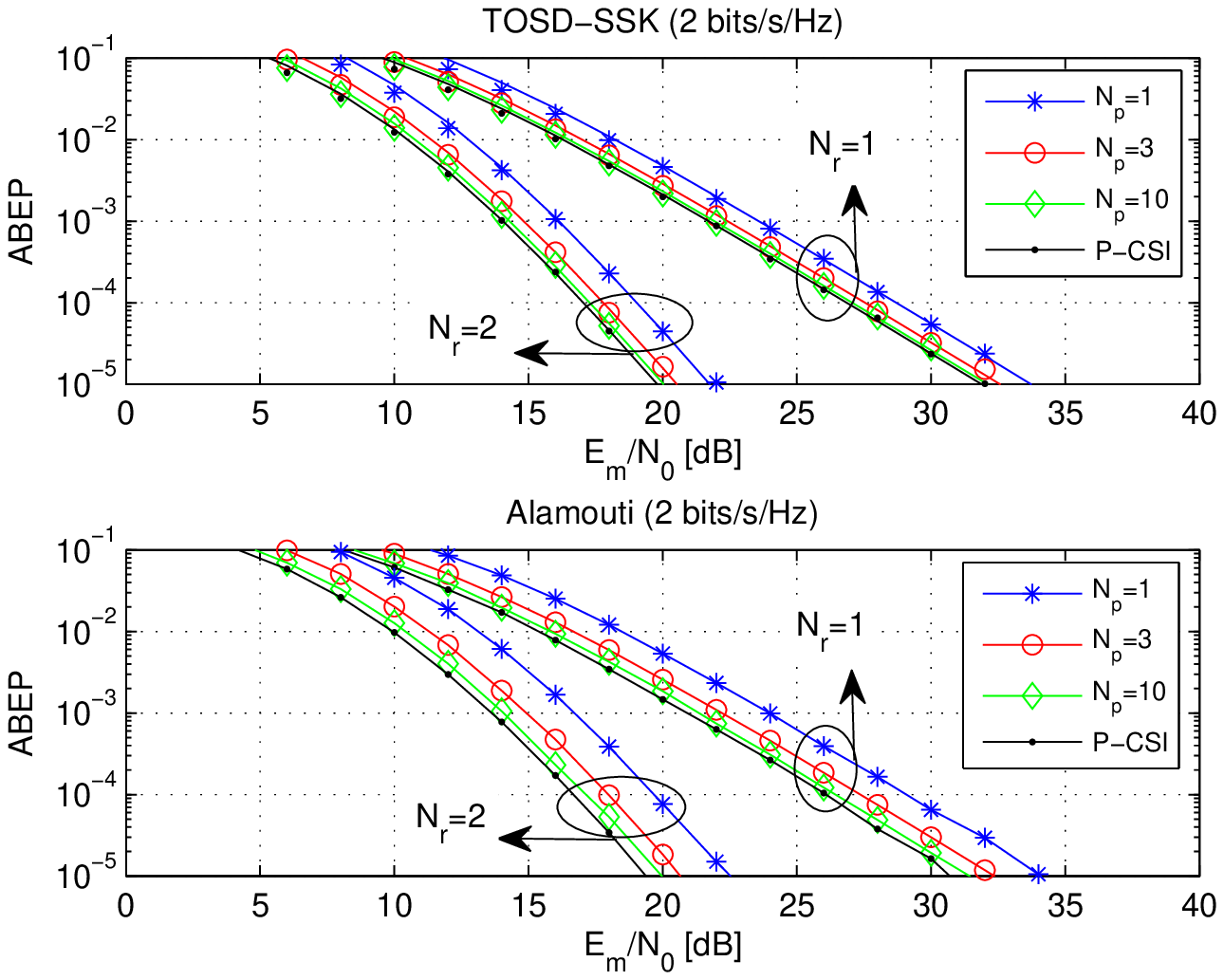}
\vspace{-0.5cm} \caption{\scriptsize ABEP against $E_m/N_0$ for various pilot pulses $N_p$ and rate 2 bits/s/Hz. (top) TOSD--SSK modulation:
solid lines show the analytical model and markers Monte Carlo simulations. (bottom) Alamouti scheme: only Monte Carlo simulations are shown.}
\label{Fig_R2}
\end{figure}

\begin{figure}[!t]
\centering
\includegraphics [width=1.1\columnwidth] {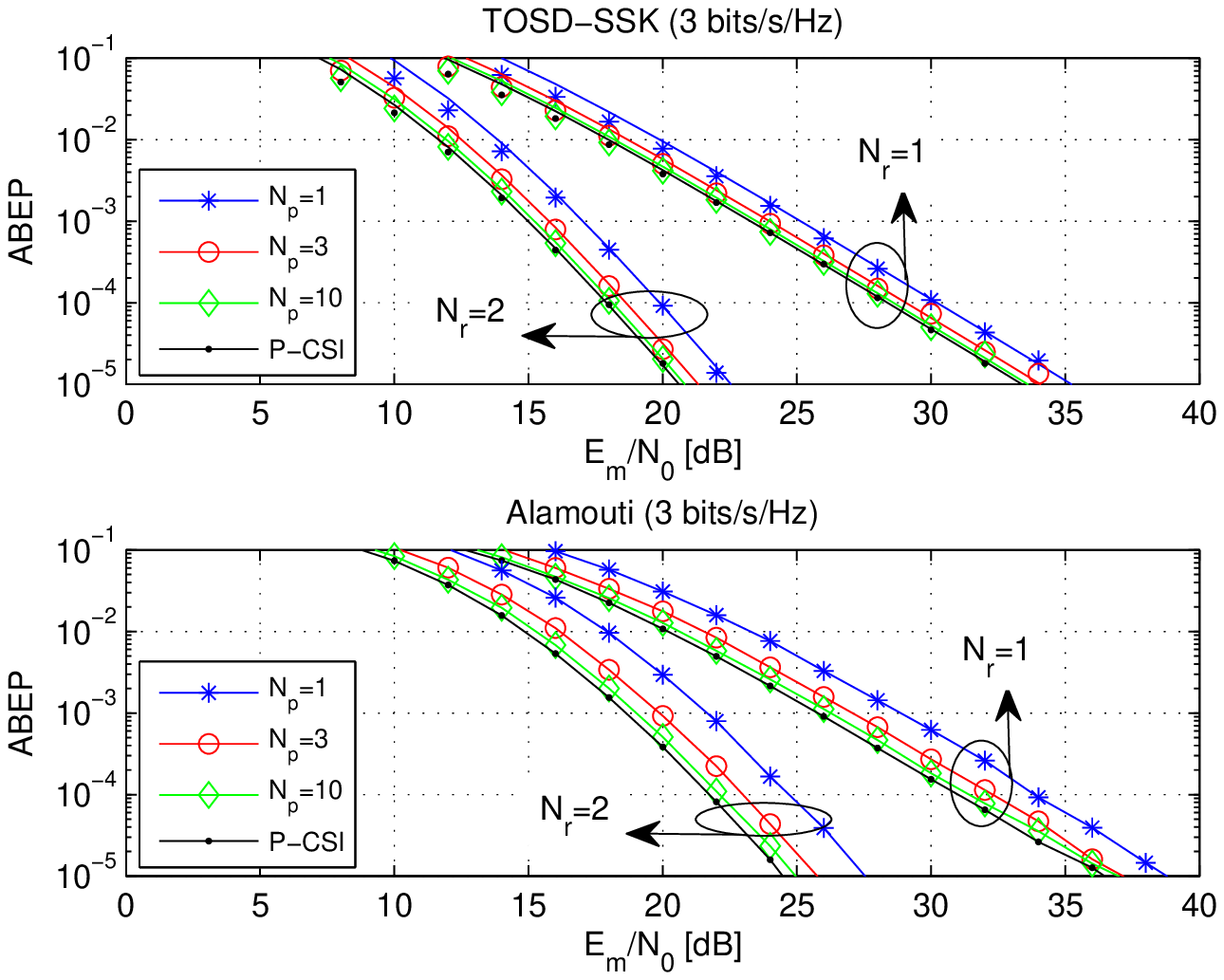}
\vspace{-0.5cm} \caption{\scriptsize ABEP against $E_m/N_0$ for various pilot pulses $N_p$ and rate 3 bits/s/Hz. (top) TOSD--SSK modulation:
solid lines show the analytical model and markers Monte Carlo simulations. (bottom) Alamouti scheme: only Monte Carlo simulations are shown.}
\label{Fig_R3}
\end{figure}

\begin{figure}[!t]
\centering
\includegraphics [width=1.1\columnwidth] {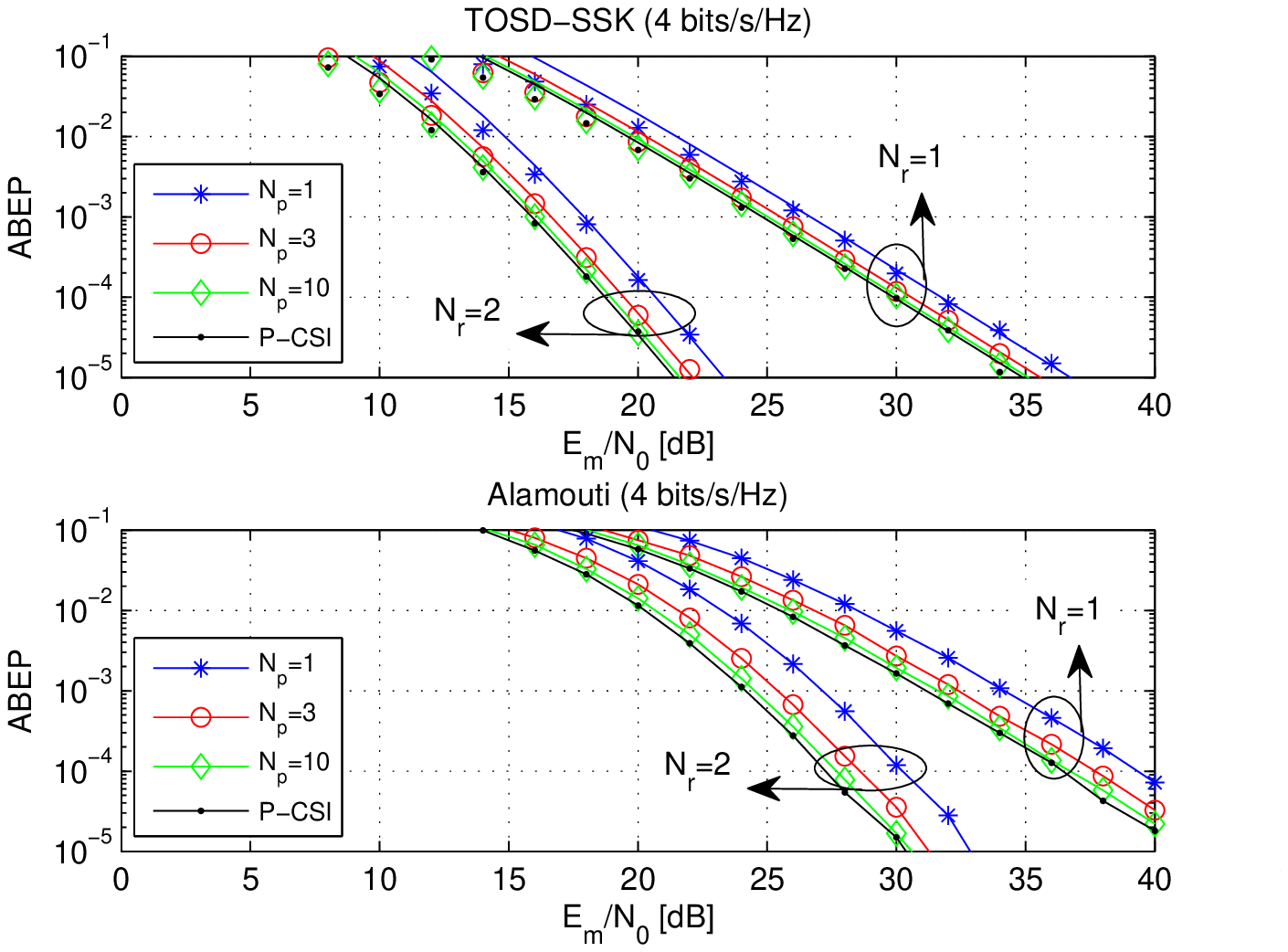}
\vspace{-0.5cm} \caption{\scriptsize ABEP against $E_m/N_0$ for various pilot pulses $N_p$ and rate 4 bits/s/Hz. (top) TOSD--SSK modulation:
solid lines show the analytical model and markers Monte Carlo simulations. (bottom) Alamouti scheme: only Monte Carlo simulations are shown.}
\label{Fig_R4}
\end{figure}

From Figs. \ref{Fig_R1}--\ref{Fig_R4}, we can observe, for various data rates, a very good agreement between Monte Carlo simulations and the
analytical framework developed in Section \ref{Performance_Analysis}. By carefully analyzing these figures, the following conclusions can be
drawn: i) TOSD--SSK modulation is quite robust to channel estimation errors and only a limited number of pilots $N_p$ are needed to get
performance very close to the P--CSI lower--bound. In particular, we notice that, in the analyzed scenario, with $N_p  = 10$ there is almost no
performance penalty; ii) the numerical examples confirm the diversity gain predicted in Section \ref{APEP}, and we notice a steeper slope when
increasing the number of antennas at the receiver. Furthermore, the diversity gain is preserved for any number of pilot pulses, and, so, the
quality of the channel estimates; iii) the performance of both TOSD--SSK modulation and Alamouti scheme gets worse for increasing values of the
data rate, as expected; iv) TOSD--SSK modulation is worse than the Alamouti scheme when the data rate is low (1 bits/s/Hz), it yields
comparable performance for medium data rates (2 bits/s/Hz), while it noticeably outperforms the Alamouti scheme for high data rate (3 bits/s/Hz
and 4 bits/s/Hz), while still keeping almost the same computational complexity. This confirms that TOSD--SSK modulation is a good candidate for
high data rate applications, and where the data rate can be increased by adding more antennas at the transmitter but still keeping only one of
them active at any time instance; and v) the TOSD--SSK modulation scheme is more robust to channel estimation errors than the Alamouti scheme,
which agrees with the results obtained in \cite{Ulla_Faiz} where SM is compared to the V--BLAST scheme. This result is very important and
breaks with the misconception that the space modulation concept is inherently less robust to inaccuracies in the channel estimation because it
maps data information on the impulse response of the wireless channel.

More specifically, the relative robustness of TOSD--SSK modulation with respect to the Alamouti scheme can be quantitatively analyzed in Table
\ref{Tab_1}, where we have (approximately) computed the SNRs needed to get ${\rm{ABEP}} = 10^{ - 4}$. For example, we notice that for TOSD--SSK
modulation the SNR gap between the setups with P--CSI and $N_p=1$ is approximately equal to 2dB, while in the same scenario the SNR penalty for
the Alamouti scheme is approximately 3dB. Furthermore, the higher robustness of TOSD--SSK modulation to imperfect channel knowledge can be
observed by carefully analyzing Table \ref{Tab_1} for a rate equal to 2 bits/s/Hz. We observe that if channel estimation is perfect, the
Alamouti scheme is slightly better than TOSD--SSK modulation. However, TOSD--SSK modulation gets slightly better when $N_p=1$. This example
clearly shows the potential benefits of TOSD--SSK modulation in practical scenarios with imperfect channel knowledge.
\section{Conclusion} \label{Conclusion}
In this paper, we have analyzed the performance of the space modulation principle when CSI is not perfectly known at the receiver. A very
accurate analytical framework has been proposed and it has been shown that, unlike common belief, transmission systems based on the space
modulation concept can be more robust to channel estimation errors than conventional modulation schemes. Furthermore, it has been shown that
only few pilot symbols are needed to achieve almost the same performance as the reference scenario with perfect channel knowledge. These
results clearly point out the usefulness of the space modulation principle in practical operating conditions, as well as that it can be a
promising low--complexity transmission technology for the next generation MIMO wireless systems.
\section*{Acknowledgment} \footnotesize
We gratefully acknowledge support from the European Union (PITN--GA--2010--264759, GREENET project) for this work. Marco Di Renzo acknowledges
support of the Laboratory of Signals and Systems (L2S) under the research project ``Jeunes Chercheurs''. Dario De Leonardis and Fabio Graziosi
acknowledge the Italian Inter--University Consortium for Telecommunications (CNIT) under the research grant ``Space Modulation for MIMO
Systems''. Harald Haas acknowledges the EPSRC (EP/G011788/1) and the Scottish Funding Council support of his position within the Edinburgh
Research Partnership in Engineering and Mathematics between the University of Edinburgh and Heriot Watt University.
\begin{table*}[!t] \scriptsize
\renewcommand{\arraystretch}{1.1}
\caption{\scriptsize Required $E_m/N_0$ (${\rm{dB}}$) to get ${\rm{ABEP}}= 10^{-4}$. The first and second lines show the setup with $N_r=1$ and
$N_r=2$, respectively. \vspace{-0.35cm}} \label{Tab_1}
\begin{center}
\begin{tabular}{|c||c|c|c|c||c||c|c|c|c|}
\hline
\multicolumn{5}{|c|} {TOSD--SSK} & \multicolumn{5}{|c|} {Alamouti}\\
\hline \hline
Rate & $N_p=1$ & $N_p=3$ & $N_p=10$ & ${\rm{P-CSI}}$ & Rate & $N_p=1$ & $N_p=3$ & $N_p=10$ & ${\rm{P-CSI}}$\\
\hline \hline
1 bits/s/Hz & $\begin{array}{*{20}c} {27.1}  \\ {18.2}  \\ \end{array}$ & $\begin{array}{*{20}c} {26}  \\ {16.9}  \\ \end{array}$ & $\begin{array}{*{20}c} {25.5}  \\ {16.4}  \\ \end{array}$ & $\begin{array}{*{20}c} {25.3}  \\ {16.2}  \\ \end{array}$ & 1 bits/s/Hz & $\begin{array}{*{20}c} {25.3}  \\ {16.2}  \\ \end{array}$ & $\begin{array}{*{20}c} {23.5}  \\ {14.5}  \\ \end{array}$ & $\begin{array}{*{20}c} {22.8}  \\ {13.5}  \\ \end{array}$ & $\begin{array}{*{20}c} {22.3}  \\ {13.2}  \\ \end{array}$\\
\hline
2 bits/s/Hz & $\begin{array}{*{20}c} {28.7}  \\ {19}  \\ \end{array}$ & $\begin{array}{*{20}c} {27.5}  \\ {17.8}  \\ \end{array}$ & $\begin{array}{*{20}c} {27}  \\ {17.3}  \\ \end{array}$ & $\begin{array}{*{20}c} {26.8}  \\ {17}  \\ \end{array}$ & 2 bits/s/Hz & $\begin{array}{*{20}c} {29.1}  \\ {19.7}  \\ \end{array}$ & $\begin{array}{*{20}c} {27.4}  \\ {18}  \\ \end{array}$ & $\begin{array}{*{20}c} {26.4}  \\ {17.1}  \\ \end{array}$ & $\begin{array}{*{20}c} {26.1}  \\ {16.7}  \\ \end{array}$\\
\hline
3 bits/s/Hz & $\begin{array}{*{20}c} {30.2}  \\ {19.8}  \\ \end{array}$ & $\begin{array}{*{20}c} {29}  \\ {18.6}  \\ \end{array}$ & $\begin{array}{*{20}c} {28.5}  \\ {18.2}  \\ \end{array}$ & $\begin{array}{*{20}c} {28.4}  \\ {17.8}  \\ \end{array}$ & 3 bits/s/Hz & $\begin{array}{*{20}c} {33.8}  \\ {24.7}  \\ \end{array}$ & $\begin{array}{*{20}c} {32.3}  \\ {23}  \\ \end{array}$ & $\begin{array}{*{20}c} {31.4}  \\ {22.1}  \\ \end{array}$ & $\begin{array}{*{20}c} {30.8}  \\ {21.7}  \\ \end{array}$\\
\hline
4 bits/s/Hz & $\begin{array}{*{20}c} {31.7}  \\ {20.7}  \\ \end{array}$ & $\begin{array}{*{20}c} {30.5}  \\ {19.4}  \\ \end{array}$ & $\begin{array}{*{20}c} {30.1}  \\ {18.9}  \\ \end{array}$ & $\begin{array}{*{20}c} {29.9}  \\ {18.7}  \\ \end{array}$ & 4 bits/s/Hz & $\begin{array}{*{20}c} {39.4}  \\ {30.2}  \\ \end{array}$ & $\begin{array}{*{20}c} {37.7}  \\ {28.6}  \\ \end{array}$ & $\begin{array}{*{20}c} {36.7}  \\ {27.7}  \\ \end{array}$ & $\begin{array}{*{20}c} {36.4}  \\ {27.2}  \\ \end{array}$\\
\hline
\end{tabular}
\end{center} \vspace{-0.5cm}
\end{table*}

\begin{thebibliography}{99} \scriptsize
%
\bibitem{Yang_2008} Y. Yang and B. Jiao,
              ``Information--guided channel--hopping for high data rate wireless communication'',
             \emph{IEEE Commun. Lett.},
             vol. 12, pp. 225--227, Apr. 2008.
%
\bibitem{Haas_TVT} R. Y. Mesleh, H. Haas, S. Sinanovic, C. W. Ahn, and S. Yun,
              ``Spatial modulation'',
             \emph{IEEE Trans. Veh. Technol.},
             vol. 57, no. 4, pp. 2228--2241, July 2008.
%
\bibitem{Ghrayeb_TWC} J. Jeganathan, A. Ghrayeb, L. Szczecinski, and A. Ceron,
              ``Space shift keying modulation for MIMO channels'',
             \emph{IEEE Trans. Wireless Commun.},
             vol. 8, no. 7, pp. 3692--3703, July 2009.
%
\bibitem{MDR_TCOM2010} M. Di Renzo and H. Haas,
              ``A general framework for performance analysis of space shift keying (SSK) modulation for MISO correlated Nakagami--\emph{m} fading channels'',
             \emph{IEEE Trans. Commun.},
             vol. 58, no. 9, pp. 2590--2603, Sep. 2010.
%
\bibitem{MDR_TOSD} M. Di Renzo and H. Haas,
              ``Performance comparison of different spatial modulation schemes in correlated fading channels'',
             \emph{IEEE Int. Conf. Commun.},
             pp. 1--6, May 2010.
%
\bibitem{MDR_TCOM_TOSD} M. Di Renzo and H. Haas,
              ``Space shift keying (SSK--) MIMO over correlated Rician fading channels: Performance analysis and a new method for transmit--diversity'',
             \emph{IEEE Trans. Commun.},
             vol. 59, no. 1, pp. 116--129, Jan. 2011.
%
\bibitem{MDR_TOSD_GSSK} M. Di Renzo and H. Haas,
              ``Space shift keying (SSK) modulation: On the transmit--diversity/multiplexing trade--off'',
             \emph{IEEE Int. Conf. Commun.},
             pp. 1--6, June 2011.
%
\bibitem{MDR_TCOM_PCSI} M. Di Renzo and H. Haas,
              ``Space shift keying (SSK) modulation with partial channel state information: Optimal detector and performance analysis over fading channels'',
             \emph{IEEE Trans. Commun.},
             vol. 58, no. 11, pp. 3196--3210, Nov. 2010.
%
\bibitem{Ulla_Faiz} M. M. Ulla Faiz, S. Al--Ghadhban, and A. Zerguine,
              ``Recursive least--squares adaptive channel estimation for spatial modulation systems'',
             \emph{IEEE Malaysia Int. Conf. Commun.},
             pp. 1--4, Dec. 2009.
%
\bibitem{V_BLAST} P. Wolniansky, G. Foschini, G. Golden, and R. Valenzuela,
              ``V--BLAST: An architecture for realizing very high data rates over the rich--scattering wireless channel'',
             \emph{IEEE Int. Symp. Signals, Systems, Electr.},
             pp. 295--300, Sep./Oct. 1998.
%
\bibitem{Biglieri} G. Taricco and E. Biglieri,
              ``Space--time decoding with imperfect channel estimation'',
             \emph{IEEE Trans. Wireless Commun.},
             vol. 4, no. 4, pp. 1874--1888, July 2005.
%
\bibitem{MDR_QF} M. Di Renzo F. Graziosi, and F. Santucci,
              ``On the cumulative distribution function of quadratic--form receivers over generalized fading channels with tone interference'',
             \emph{IEEE Trans. Commun.},
             vol. 57, no. 7, pp. 2122--2137, July 2009.
%
\bibitem{Alamouti} S. M. Alamouti,
              ``A simple transmit diversity technique for wireless communications'',
             \emph{IEEE J. Sel. Areas Commun.},
             vol. 16, no. 8, pp. 1451--1458, Oct. 1998.
%
\bibitem{TCOM_PSM} J. A. Ney da Silva and M. L. R. de Campos,
              ``Spectrally efficient UWB pulse shaping with application in orthogonal PSM'',
             \emph{IEEE Trans. Commun.},
             vol. 55, no. 2, pp. 313--322, Feb. 2007.
%
\bibitem{TD_CommMag} R. Derryberry, {\emph{et al.}},
              ``Transmit diversity in 3G CDMA systems'',
             \emph{IEEE Commun. Mag.},
             vol. 40, vol. 4, pp. 68-75, Apr. 2002.
%
\bibitem{Proakis_1968} J. G. Proakis,
              ``Probabilities of error for adaptive reception of M--phase signals'',
             \emph{IEEE Trans. Commun. Technol.},
             vol. COM--16, no. 1, pp. 71--81, Feb. 1968.
%
\bibitem{Gifford_2005} W. M. Gifford, M. Z. Win, M. Chiani,
              ``Diversity with practical channel estimation'',
             \emph{IEEE Trans. Wireless Commun.},
             vol. 4, no. 4, pp. 1935--1947, July 2005.
%
\bibitem{Simon} M. K. Simon and M.--S. Alouini,
              \emph{Digital Communication over Fading Channels},
             John Wiley $\&$ Sons, Inc., 1st ed., 2000.
%
\bibitem{VanTrees} H. L. Van Trees,
             \emph{Detection, Estimation, and Modulation Theory, Part I: Detection, Estimation, and Linear Modulation Theory},
             John Wiley $\&$ Sons, Inc. 2001.
%
\bibitem{Giannakis} Z. Wang and G. B. Giannakis,
              ``A simple and general parameterization quantifying performance in fading channels'',
             \emph{IEEE Trans. Commun.},
             vol. 51, no. 8, pp. 1389--1398, Aug. 2003.
%
%
%
\end{thebibliography}
\end{document}